# Predicted response of an atom to a short burst of electro-magnetic radiation

V.V. Semak[1,*] and M.N.Shneider[2,#]

[1]SARA Inc., Cypress CA

[2]Princeton University, Princeton NJ

[*]vsemak@sara.com,   [#]m.n.shneider@gmail.com

**Abstract**

In this work we present a hypothesis that spectral response of atoms or molecules to a pulse of electro-magnetic radiation with fast rising or falling fronts would contain a unique emission line that is located approximately near the frequency of the natural oscillations of optical electron. The emission of this "pinging" spectral line would exist during the time that is determined by the time of radiative energy loss by the optical electron. The amplitude of the "pinging" spectral line would be higher for the pulses with faster raising or falling fronts. The simulations using our previously developed model confirmed existence of the "pinging" spectral response. If experimentally confirmed, this work could lead to a new high sensitivity and high signal to noise ratio stand-off detection techniques, a new type of LIDAR, and to other yet unknown applications.

**Introduction**

Current theoretical model of how an atom responds to an electro-magnetic wave is based on the well-known Lorentz Oscillator Model [1]. It is commonly stated (for example [2, 3, 4]) that this model is surprisingly accurate in spite its simplicity. Driven by curiosity at first, we studied the origins of the Lorentz's model and evaluated its predictive capability. This study unveiled that Lorentz Oscillator Model is based on several assumptions that could be traced back to the believes of physics of the late 19[th] century. These believes are in contradiction with the discoveries and models of the modern science. A reader could find some of our analysis in the Arxiv submission [5]. There, a reader would also find our new model of atomic response to the electro-magnetic radiation that is consistent with the modern physical models. A numerical tool recently developed on the base of our new model demonstrated to accurately predict effects observed in nonlinear optics experiments [6] that is beyond the capability of the Lorentz model. Additionally, the new theoretical concept [5] leads to a suggestion of existence of an effect yet unobserved in optics. Our model predicts that an electron, bound in an atom and exposed to a pulse of electro-magnetic radiation, would undulate re-emitting at the frequency of the incident light, $\omega_L$ and at the multiple harmonics of this frequency. Besides that, it will emit at the frequency that is close to what is typically called "the natural oscillation frequency", $\omega_0$. The former re-emission is well known as scattering, higher harmonic generation and optical parametric oscillation. However, the latter emission is yet unknown. Thus, we predict that "pinging" an atom with a fast front electro-magnetic pulse will produce emission of yet unobserved spectral line that would have central frequency determined by the frequency of the electron natural oscillations in the atom and the time constant of the radiative decay of the oscillation energy of the optical electron. Here, we should mention that this effect is well known in the theory of mechanical and electric oscillators [7]; however, in optics the nonstationary (transient) behavior of the oscillations of a bound electron was never consider prior to our work. If experimentally confirmed, the technique of "pining" spectral response of atoms and molecules would facilitate development of multiple technologies.

**Theoretical Model**

According to our theoretical model [5] a general form of equation describing response of an electron bound in an atom or molecule to electro-magnetic radiation is given by the following equation:

$$\ddot{r} + \frac{1}{m_e}\frac{\partial U(r)}{\partial r} - \frac{\xi}{m_e}\dddot{r} - \frac{1}{m_e}F_{mb}(r) = -\frac{e}{m_e}E(t), \qquad (1)$$

where $m_e$ is the electron mass, $r$ is the radial position of electron in the atom or molecule, $U(r)$ is the potential in which electron resides, $\xi = e^2/6\pi\varepsilon_0 c^3$, $F_{mb}(r)$ is the force due to nonradiative loss of electron energy, $e$ is the electron charge, and $E$ is the electric field of the incident electro-magnetic wave.

The first term in the left hand side is the acceleration of optical electron induced by the action of the electro-magnetic wave that is second derivative of the radial position of the electron.

The electron is bound and resides in a potential $U(r)$, Therefore, the displacement from the equilibrium position would produce a "returning" force that equals to the gradient of the potential. Thus, the second term in the left hand side of the equation (1) describes the component of electron acceleration caused by this returning force.

According to the postulates of quantum mechanics, an electron residing on the equilibrium orbit does not emit radiation although it moves with acceleration. However, we suggest that, if the electron is displaced from the equilibrium orbit and moves with acceleration, it emits radiation. Note, that at this point we are not considering the case when the external force can induce electron transition from one stable orbit to another stable orbit. Thus, the absorption or emission of radiation associated with the transitions between the energy levels is not included in this consideration. Instead, we consider a "small" action that undulates electron near its stable orbit and is insufficient to induce quantum mechanical transition between the orbits. Such electron undulations should produce radiation since it is not banned by the mentioned postulate. This radiation is what is known as scattering. This forced undulation also produces harmonics and sub-harmonics. It has been shown that the damping force associated with the energy loss due to the dipole radiation is proportional to the third order time derivative of the electron displacement [8] that gives the third term in the left hand side of the equation (1).

It is reasonable to suggest that, there are possibly some other mechanisms that resist motion of the electron displacement from the equilibrium orbit that are non-radiative. Our thorough research demonstrated that, there is no concept of possible mechanisms for the nonradiative dissipation of electron energy associated with externally induced undulations near its equilibrium position. Therefore, in our current consideration we include deceleration force due to only radiative energy loss leaving the nonradiative mechanisms for the future research efforts.

Expressing the second term in the left hand side of the equation (1), i.e. term describing the "returning" force, is a non-trivial task because it requires description of the potential in which electron resides in an atom. As we concluded after detail studies of the current state of the art, there is no consensus in quantum mechanical description of an atom that, possibly, explains difficulties in application of these models to the description of the optical properties of matter.

In optics the atomic oscillator is described by the classical models that are based on the well-known Lorentz oscillator model. In these models the electron potential is assumed to be a quadratic function of electron displacement from equilibrium. Thus, as all optics textbooks suggest, the electron potential has the following form:

$$U(r) \sim (r - r_0)^2, \tag{2}$$

where $r_0$ is an equilibrium radial position of the electron. As a result, all current Lorentz-like atomic oscillator models treat the displacement resisting force is a linear function of the electron displacement. It is worth stating that this theoretical concept is inconsistent with or contradicts to all the current concepts of how atom is designed.

In the recent DARPA supported project we have created a numerical tool [6] that embodied our theoretical model proposed earlier [5]. During this development project, we have discovered that, the description of the electron potential based on the Bohr's model of atom, provides exceptional accuracy of prediction of the nonlinear optical effects using as input the fundamental properties of mater. In this model of atomic oscillator [5], the motion of electron on a s-orbit of a hydrogen like atom forced by the electro-magnetic wave is described as follows:

$$\ddot{r} + \frac{2U_0 r_0^2}{m_e}\left(\frac{1}{r_0 r^2} - \frac{1}{r^3}\right) - \frac{\xi}{m_e}\dddot{r} = -\frac{e}{m_e}E(t), \tag{3}$$

where $r_0$ is the equilibrium radius of the orbit and $U_0$ is the depth of effective electron potential, i.e. the ionization potential of the atom.

**Results of the numerical experiments**

We applied our recently developed theory of atomic oscillator [5] in order to verify our hypothesis that the re-emission produced by hydrogen atom in response to a pulse of electro-magnetic radiation with rapidly raising front would contain the radiation line that is close to the natural oscillation frequency of the optical electron. In these simulations we computed forced electron motion and resulting dipole radiation and, then, analyzed the spectrum of the re-emitted light. The hydrogen was modeled as a Bohr's atom with electron in s-state with the ionization potential $U_0$=13.6 eV. The frequency of natural oscillations of the electron is approximately $\omega_0 = \left(\frac{2U_0}{mr_0^2}\right)^{1/2}$; $\nu_0 = \frac{\omega_0}{2\pi} = 6.57 \cdot 10^{15}$ Hz.

The computations were performed for almost rectangular temporal shape the laser pulse with duration $\tau_L = 10$ ns. The laser pulse has fast raising front reaching constant laser intensity:

$$I(t) = I_L\left(1 - \exp(-t/t_r)\right)^2, \tag{4}$$

where the characteristic rise time was chosen to be 100 periods of laser oscillations, i.e. $t_r = 100\left(\frac{\omega_L}{2\pi}\right)^{-1}$. The spatial laser beam intensity distribution was assumed to be uniform with radius $r_L = 20$ μm. The computations were performed for two laser pulse energies $E_L = 10$ mJ and $E_L = 0.01$ mJ with corresponding intensities $I_L = E_L/\pi r_L^2 \tau_L$=7.95·$10^{14}$ W/m² and $I_L = 7.95 \cdot 10^{11}$ W/m².

The computation time for each case was chosen to be first 5000 laser oscillation periods after which the spectrum of the forced dipole radiation was computed. Note, that the time for the radiation decay due to the dipole emission exceeds the time of 5000 laser cycles, i.e. the computed spectra correspond to the transient regime of atomic oscillator.

We computed the power spectral densities of radiation re-emitted by an atom for higher and lower laser intensities at the 532nm laser wavelength (Figure 1 and 2) and for lower laser intensity at the 266nm laser wavelength (Figure 3) and for higher intensity at the 1064nm laser wavelength (Figure 4). Note that both higher and lower intensities were insufficient to produce noticeable nonlinear response, i.e.

harmonic generation. Hence, although the harmonics were present, their intensities were too miniscule to appear on the linear scale used for plotting the spectra of electron re-emission.

The simulations confirmed our hypothesis, that "pinging" an atom with an electro-magnetic pulse with sharp front would produce emission at the frequency close to the natural oscillation frequency. The results presented in the Figures 1-3 demonstrate re-emission at the frequency of the laser and the emission of noticeable strength at the frequency that is close to $\nu_0 = \frac{\omega_0}{2\pi} = 6.57 \cdot 10^{15}$ Hz. The "pinging" frequency is independent of the laser frequency since it is determined by the fundamental atomic properties, namely, ionization potential, radius of the equilibrium orbit of the electron, and the radiative decay time. Note, that in case of hydrogen, the "pining" line is expected to be in the soft X-ray spectral range.

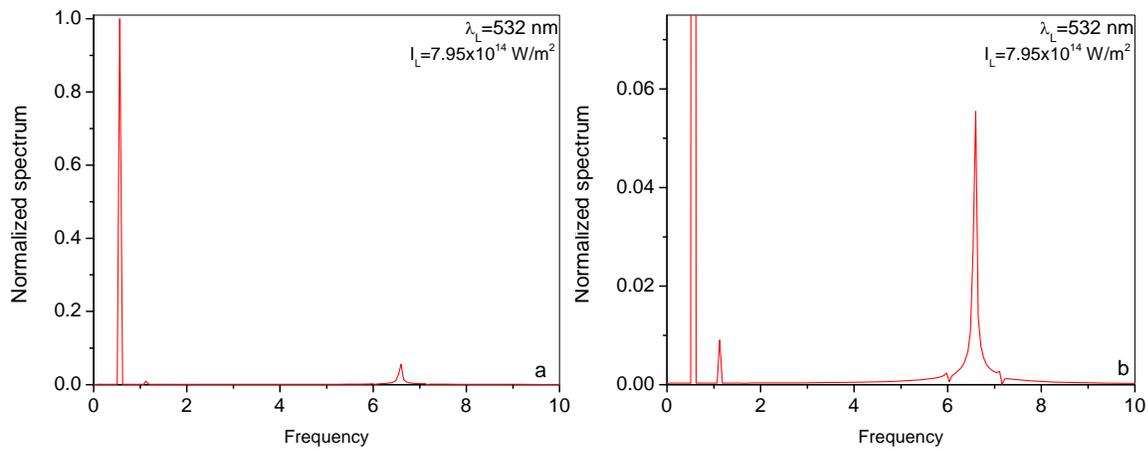

**Figure 1**. a) Spectral response of hydrogen to irradiation with 10 mJ, 10 ns laser pulse at 532 nm wavelength, focused in the spot with uniform intensity distribution with 20 m radius. The x-axis represents the frequency and has dimension of $10^{15}$ Hz. b) – the vertical scale is magnified to show the 2nd harmonic and "pinging" lines.

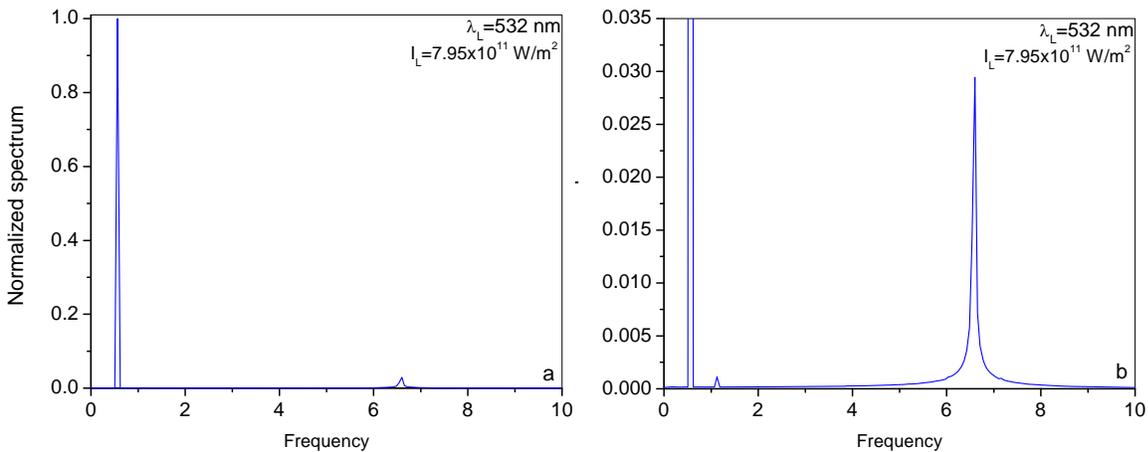

**Figure 2.** a) Spectral response of hydrogen to irradiation with 0.01 mJ, 10 ns laser pulse at 532 nm wavelength, focused in the spot with uniform intensity distribution with 20 m radius. The x-axis represents the frequency and has dimension $10^{15}$ Hz. b) same as previous.

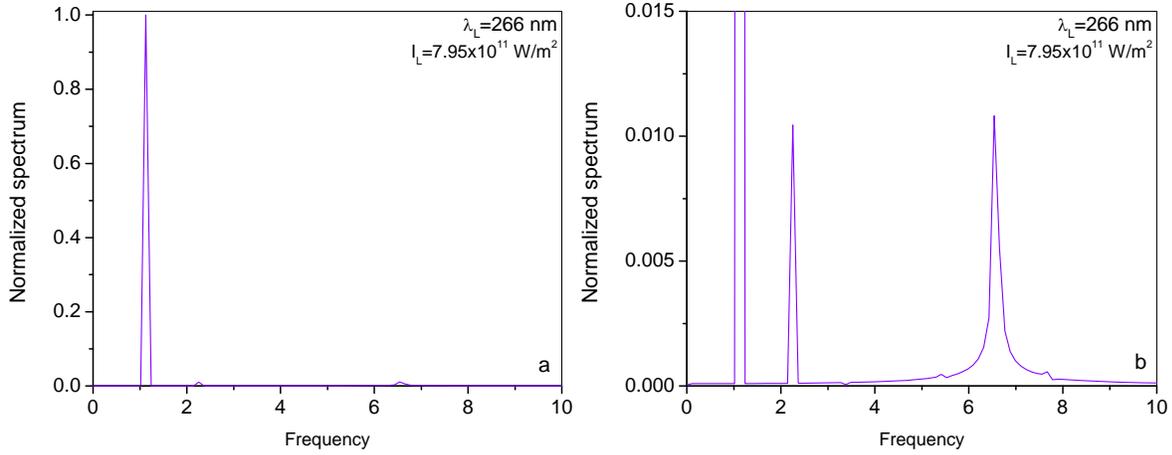

**Figure 3.** a) Spectral response of hydrogen to irradiation with 0.01 mJ, 10 ns laser pulse at 266 nm wavelength, focused in the spot with uniform intensity distribution with 20 m radius. The x-axis represents the frequency and has dimension $10^{15}$ Hz. b) same as previous.

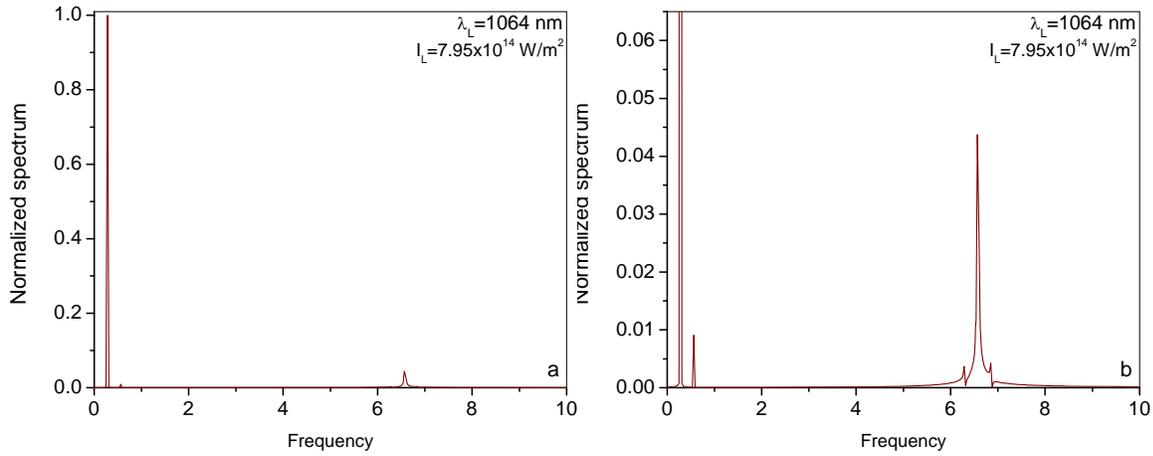

**Figure 4.** a) Spectral response of hydrogen to irradiation with 10 mJ, 10 ns laser pulse at 1064 nm wavelength, focused in the spot with uniform intensity distribution with 20 m radius. The x-axis represents the frequency and has dimension $10^{15}$ Hz. b) same as previous.

Note that, the relative amplitudes of the lines in the power spectral density of the atomic re-radiation at the frequencies of the incident radiation and the "pinning" frequency can be controlled by varying the parameters of the incident pulse of electro-magnetic radiation.

## Linear vs non-linear oscillator

It is important to point out that the emission of "pinging" spectral line by an atom in response to a sharp pule isn't a unique property of a nonlinear optical oscillator, i.e. oscillator with a nonquadratic potential $U(r)$ and radiation damping proportional to the third order time derivative of the electron displacement, $\propto \dddot{r}$. Even in the case if atomic oscillator is assumed to be linear oscillator, as assumed by the current models based on classical Lorentz oscillator, the "pinging" re-emission would be also predicted by such a model. Our simulations of the optical response of linear Lorentz oscillator verifies presence of the "pinging" line in the re-emission spectra. As the calculations show (Fig. 5, a, b), the difference between the response of a nonlinear and a linear atomic oscillator to a laser pulse with fast raising front is only that the 2nd harmonic that appears in the nonlinear oscillator simulations absent for the Lorentz oscillator. In the same time, the "pining" spectral line is present in both simulations.

The re-emission at the frequency close to the natural frequency of atomic oscillator would be detectable if the characteristic decay time of this radiation is longer than the period of the natural oscillations, $\tau_\gamma > 1/\omega_0$. If an atom is exposed to a long pulse of harmonic field oscillating with frequency $\omega_L$, it will re-emit radiation at the forcing frequency $\omega_L$, at the frequencies of harmonics and (possibly) sub-harmonics, and at the "pinging" frequency. Note that the "pinging" emission would take place only during the transient stage, i.e. it will be present in the re-emission spectrum during time determined by the radiative decay time, $\tau_\gamma$. Moreover, the amplitude of emission at the "pinging" frequency would be greater for the sharper fronts of the driving force.

This can be understood from the solution for a linear oscillator with harmonic external force [7,9]:

$$\ddot{r} + \gamma \dot{r} + \omega_0^2 r = -\frac{eE_0}{m_e} e^{-i\omega t}, \tag{5}$$

where $\gamma = \frac{\xi}{m_e}\omega_0^2 = \frac{e^2 \omega_0^2}{6\pi\varepsilon_0 m_e c^3}$ [1/sec]. Note, that for the hydrogen atom under consideration, the characteristic damping time of electron natural oscillations is $\tau_\gamma \sim 7.5 \cdot 10^{-10}$ sec. This is more than six orders of magnitude larger than the period of natural oscillations, $T_{\omega_0} = \frac{2\pi}{\omega_0} \approx 1.52 \cdot 10^{-16}$ sec.

For the initial conditions

$$r(0) = 0, \quad \dot{r}(0) = 0 \tag{6}$$

in case of slowly decaying natural oscillations $\gamma \ll 2\omega_0$, the general solution of equation (5) with the initial conditions (6) gives the real part of complex displacement $r(t)$, [7]:

$$r = \frac{eE_0}{m_e}\frac{(\omega^2-\omega_0^2)\cos(\omega t)-\omega\gamma\sin(\omega t)}{(\omega^2-\omega_0^2)^2+\omega^2\gamma^2} - \frac{eE_0}{m_e}e^{-\frac{1}{2}\gamma t}\frac{(\omega^2-\omega_0^2)\cos(\omega' t)-\frac{(\omega^2+\omega_0^2)}{2\omega'}\gamma\sin(\omega' t)}{(\omega^2-\omega_0^2)^2+\omega^2\gamma^2}, \tag{7}$$

where $\omega' = \left(\omega_0^2 - \frac{1}{4}\gamma^2\right)^{1/2}$ is the "pinging" frequency that, for small $\gamma \ll \omega_0$, is very close to the natural oscillation frequency $\omega_0$.

If the forcing frequency is far from the resonance and assuming that the damping of natural oscillations is slow, i.e., when

$$\omega\gamma \ll \omega^2 - \omega_0^2; \quad \omega, \omega_0 \gg \gamma,$$

from (7) we find the frequency dependence of the field of the reradiation of the incident wave

$$E(t,\omega) \propto \ddot{r}(t,\omega) \approx -\frac{eE_0}{m_e}\frac{\omega^2\cos(\omega t)}{\omega^2-\omega_0^2} + \frac{eE_0}{m_e}e^{-\frac{1}{2}\gamma t}\frac{\omega'^2\cos(\omega' t)}{\omega^2-\omega_0^2}. \tag{8}$$

The equation (8) proves that, for a linear oscillator and elastic (Rayleigh) scattering conditions, the re-emission during the transient stage produces two spectral lines: one line is at the frequency of the forced oscillations, and the second line is at frequency that is close to the natural frequency of the electron oscillations ("pinging" spectral line).

This analytical result is obtained for the case of instantaneously increasing electric field at the moment when the electro-magnetic wave arrives to the atom. For the case of realistic laser front with the characteristic raise time of the front that is approximately 100 of laser oscillation periods we compared numerically the spectra of re-emitted light for the linear and non-linear oscillators. The simulation results clearly show that the intensity of the "pining" frequency for both oscillators is similar (Figure 5). Of course, as previously mentioned, the computations using realistic (non-quadratic) electron potential show presence of second harmonic (Figure 5 b) unlike computations assuming the linearity of the atomic oscillator.

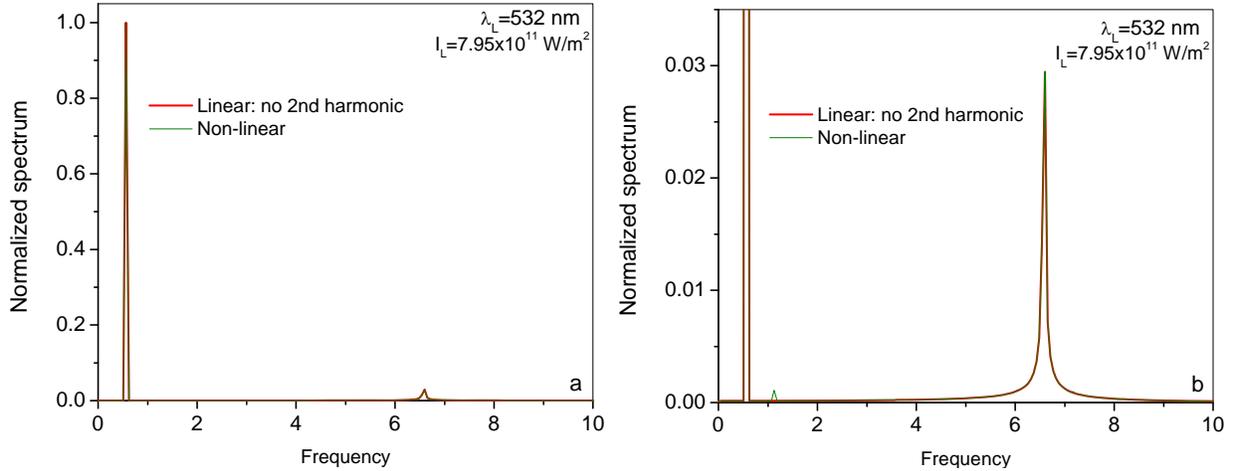

**Figure 5.** Re-emission spectra computed for nonlinear and linear oscillators showing similar generation of "pining" spectral line; a – the vertical scale is normalized to the scattered re-radiation line, b – the vertical scale is magnified to show the "pinging" line (note the second harmonic line generated by nonlinear oscillator and is absent on the spectrum of linear oscillator).

## Conclusions

The numerical simulations using our recently developed model of atomic oscillator with realistic electron potential confirmed our hypothesis that materials responding to the pulses of electro-magnetic radiation with fast fronts (like in case of ultrashort laser pulses) must contain emission of a unique line centered at a frequency that is approximately equal to the frequency of natural oscillations of optical electron. This frequency, that we call "pinging" frequency, depends only on the depth of the potential in which the optical electron resides and on the size of the electron orbit. Obviously, the shape of the electron potential and the geometry of the orbit would affect the central frequency and the shape of this "pinging" emission line. It is important to note that the "pinging" frequency is independent of the frequency of the incident electro-magnetic wave. The emission of the "pinging" spectral line would proceed during time that is determined by the time of radiative energy loss of the optical electron. The amplitude of the "pinging" spectral line would be higher for the faster raising pulse front. If experimentally confirmed, this work could lead to development of novel type of spectroscopy. In particular, the development could lead to creation of new high signal to noise ratio and high sensitivity standoff detection technologies. It is reasonable to envision development of a new type of LIDAR detection that would provide simultaneously high-resolution detection of position and velocity. The new theory of atomic response to electro-magnetic pulses will provide contribution to the theory of elastic scattering and, potentially, will open other applications.

**Acknowledgement**

V. Semak would like to express his gratitude to Dr. Ricardo A. Pakula for multiple discussions of his publications in Arxiv and intriguing insights into the current state of quantum mechanics that followed from these discussions.